\documentclass[sigconf,screen]{acmart} 
\usepackage{algorithm}
\usepackage[noend]{algpseudocode}
\usepackage{algorithmicx}
\usepackage{graphicx}
\usepackage{textcomp}
\usepackage{xcolor}

\usepackage{makecell}
\usepackage{array}
\usepackage{booktabs}

\usepackage{xcolor, soul}
\usepackage{xspace}
\usepackage{multirow}
\usepackage{tabularx}
\usepackage{tikz}
\usepackage[bottom]{footmisc}

\usepackage{fancyhdr}
\usepackage{mathtools}

\usepackage{subcaption}
\usepackage{booktabs}

\usepackage{pgfplots}
\pgfplotsset{compat=1.17} %

\usepackage{stfloats}

\usepackage{enumitem}

\usepackage{mwe}

\definecolor{lavenderr}{rgb}{0.71, 0.49, 0.86}

\usepackage{graphicx}

\usepackage{tikzducks}

\usepackage[en-GB]{datetime2} 
\DTMsetup{
    showzone=false,   
    showseconds=false 
}

\definecolor{darkspringgreen}{rgb}{0.09, 0.45, 0.27}
\definecolor{denim}{rgb}{0.08, 0.38, 0.74}
\definecolor{darkolivegreen}{rgb}{0.33, 0.42, 0.18}
\definecolor{tangerine}{rgb}{0.95, 0.52, 0.0}
\definecolor{mahogany}{rgb}{0.75, 0.25, 0.0}
\definecolor{coolblack}{rgb}{0.0, 0.22, 0.44}
\definecolor{darkpink}{rgb}{0.91, 0.35, 0.6}
\definecolor{darkblue}{rgb}{0.0, 0.0, 0.67}
\definecolor{melon}{rgb}{0.97, 0.69, 0.67}

\definecolor{seagreen}{rgb}{0.18, 0.55, 0.34}

\definecolor{pred}{rgb}{0.7843, 0.0039, 0.3137} 

\definecolor{darkpink}{rgb}{0.88, 0.28, 0.54}
\definecolor{forestgreen}{rgb}{0.0, 0.27, 0.13}
\definecolor{amber}{rgb}{1.0, 0.49, 0.0}

\newcommand{\inum}[1]{(\textit{#1})\xspace}

\newcommand{\sect}[1]{{§#1}\xspace} %
\newcommand{\head}[1]{{\vspace{2pt}\noindent\textbf{#1.}\xspace}} %
\newcommand{\fig}[1]{{Fig.~#1}\xspace} %

\newcommand\base{\textsf{Base}\xspace}

\newcommand\ssdl{\texttt{SSD-L}\xspace}

\newcommand\gs{\textsf{GS}\xspace}

\newcolumntype{Y}{>{\centering\arraybackslash}X}
\usepackage{tikz}

\newcommand{\squishlist}{
 \begin{list}{$\circ$}
  { \setlength{\itemsep}{0pt}
     \setlength{\parsep}{0pt}
     \setlength{\topsep}{3pt}
     \setlength{\partopsep}{0pt}
     \setlength{\leftmargin}{1em}
     \setlength{\labelwidth}{1em}
     \setlength{\labelsep}{0.5em} } }

\newcommand{\squishend}{
  \end{list}  }

\makeatletter
\g@addto@macro{\normalsize}{%
  \setlength{\abovedisplayskip}{0pt plus 0.5pt minus 1pt}
  \setlength{\belowdisplayskip}{4pt plus 0.5pt minus 1pt}
  \setlength{\abovedisplayshortskip}{0pt}
  \setlength{\belowdisplayshortskip}{0pt}
  \setlength{\intextsep}{2pt plus 1pt minus 1pt}
  \setlength{\textfloatsep}{5pt plus 1pt minus 1pt}
  \setlength{\skip\footins}{3pt plus 1pt minus 1pt}}
\makeatother

\usepackage{blindtext,graphicx}
\usepackage[absolute]{textpos}
\usepackage{datetime}

\definecolor{seagreen}{rgb}{0.18, 0.55, 0.34}
\definecolor{ballblue}{rgb}{0.13, 0.67, 0.8}

\newcommand\omccc[1]{{{#1}}}
\newcommand\omciv[1]{{{#1}}}

\newcommand\omcvii[1]{{{#1}}}

\newcommand\jsr[1]{{{#1}}}
 \newcommand\rev[1]{{{#1}}}

\definecolor{darkgreen}{rgb}{0.0, 0.44, 0.34}

\sloppypar

\definecolor{dollarbill}{rgb}{0.52, 0.73, 0.4}

\usetikzlibrary{patterns}
\usepackage[precision=2, unit=mm]{lengthconvert}

\usepackage[colorinlistoftodos,prependcaption,textsize=scriptsize]{todonotes}
\usepackage{marginnote} 
\definecolor{cyan(process)}{rgb}{0.0, 0.62, 0.82}

\definecolor{cadmiumgreen}{rgb}{0.0, 0.50, 0.29}

\newcommand\revref[1]{\hyperref[rev:#1]{#1}}

\definecolor{raspberry}{rgb}{0.89, 0.04, 0.36}

\definecolor{awesome}{rgb}{1.0, 0.13, 0.32}
\definecolor{cardinal}{rgb}{0.77, 0.12, 0.23}
\definecolor{cadet}{rgb}{0.33, 0.41, 0.47}
\definecolor{celadon}{rgb}{0.67, 0.88, 0.69}
\definecolor{persianblue}{rgb}{0.11, 0.22, 0.73}
\definecolor{ultramarine}{rgb}{0.07, 0.04, 0.56}
\definecolor{warmblack}{rgb}{0.0, 0.3, 0.3}

\definecolor{terracotta}{rgb}{0.89, 0.45, 0.36}

\definecolor{forestgreen(web)}{rgb}{0.13, 0.55, 0.13}

\renewcommand{\figurename}{Fig.}

\definecolor{cardinal}{rgb}{0.77, 0.12, 0.23}
\definecolor{deeppink}{rgb}{1.0, 0.08, 0.58}
\definecolor{brightpink}{rgb}{1.0, 0.0, 0.5}
\definecolor{electricviolet}{rgb}{0.56, 0.0, 1.0}
\definecolor{brandeisblue}{rgb}{0.0, 0.44, 1.0}
\definecolor{carminered}{rgb}{1.0, 0.0, 0.22}

\definecolor{acolor}{rgb}{0.0, 0.5, 1.0}
\definecolor{bcolor}{rgb}{0.54, 0.17, 0.89}
\definecolor{ccolor}{rgb}{0.4, 0.69, 0.2}
\definecolor{dcolor}{rgb}{0.92, 0.41, 0.12}
\definecolor{ecolor}{rgb}{0.6, 0.0, 0.156}
\definecolor{fcolor}{rgb}{0.106, 0.620, 0.467}

\newcommand\icsom[1]{{\color{black}{#1}}}

\newcommand\icsomm[1]{{\color{black}{#1}}}
\newcommand\icsommm[1]{{\color{black}{#1}}}

\renewcommand\todo[1]{}

\newcommand\omiii[1]{{\color{black}{#1}}}

\definecolor{dogwoodrose}{rgb}{0.84, 0.09, 0.41}

\definecolor{turquoise}{rgb}{0.19, 0.84, 0.78}
\definecolor{brightturquoise}{rgb}{0.03, 0.91, 0.87}
	\definecolor{cinnamon}{rgb}{0.82, 0.41, 0.12}

\definecolor{azure}{rgb}{0.0, 0.5, 1.0}

\definecolor{emerald}{rgb}{0.20, 0.65, 0.38}

\definecolor{dogwoodrose}{rgb}{0.84, 0.09, 0.41}

\newcommand{\citesbs}{bentley_accurate_2008,margulies_genome_2005,shendure_accurate_2005,harris_single-molecule_2008,turcatti_new_2008,wu_termination_2007,fuller_rapid_2007,mckernan_reagents_2008,fuller_method_2011}
\newcommand{\citesmrt}{eid_real-time_2009}
\newcommand{\citenanopore}{menestrina_ionic_1986,cherf_automated_2012,manrao_reading_2012,laszlo_decoding_2014,deamer_three_2016,kasianowicz_characterization_1996,meller_rapid_2000,stoddart_single-nucleotide_2009,laszlo_detection_2013,schreiber_error_2013,butler_single-molecule_2008,derrington_nanopore_2010,song_structure_1996,walker_pore-forming_1994,wescoe_nanopores_2014,lieberman_processive_2010,bezrukov_dynamics_1996,akeson_microsecond_1999,stoddart_nucleobase_2010,ashkenasy_recognizing_2005,stoddart_multiple_2010,bezrukov_current_1993,zhang_single-molecule_2024}

\newcommand{\citesequencing}{\citesbs,\citesmrt,\citenanopore,stoler2021sequencing,goodwin2016coming,davis2021sequencerr,sereika2022oxford,jain2018nanopore,payne2018bulkvis,amarasinghe2020opportunities,hon2020highly,ni2023benchmarking,wenger2019accurate}

\newcommand{\citepersonalized}{alkan_personalized_2009,lightbody_review_2019,morganti_next_2019,branco_bioinformatics_2021,quazi_artificial_2022,aronson_building_2015,f_lochel_comparative_2020,papadopoulou_application_2023,tafazoli_applying_2021,gambardella_personalized_2020,leary_development_2010,hamburg_margaret_a_path_2010,van_der_lee_technologies_2020,moon_precision_2022,mohan_profiling_2020,chung_rapid_2020,bielinski_preemptive_2014,ho_enabling_2020,hussen_emerging_2022,russell_pharmacogenomics_2021,verma_nanopore_2024,clark2019diagnosis,farnaes2018rapid,sweeney2021rapid,flores2013p4,ginsburg2009genomic,chin2011cancer,Ashley2016}

\newcommand{\citeoutbreakrapid}{dunn_squigglefilter_2021,bertelli_rapid_2013,arias_rapid_2016,comin_investigation_2020,Quick2016}

\newcommand{\citeoutbreak}{\citeoutbreakrapid,robinson_genomics_2013,fournier_clinical_2014,koser_routine_2012,eloit_diagnosis_2014,gardy_jennifer_l_whole-genome_2011,taylor_angela_j_characterization_2015,quainoo_scott_whole-genome_2017,goldberg_brittany_making_2015,besser_interpretation_2019,li_application_2021,deng_integrated_2021,kwong_whole_2015,deurenberg_application_2017,tang_infection_2017,croucher_application_2015,
bloom2021massively,yelagandula2021multiplexed,le2013selected,nikolayevskyy2016whole,qiu2015whole,gilchrist2015whole}

\newcommand{\citeagriculture}{the_arabidopsis_genome_initiative_analysis_2000,zhu_applications_2020,choi_nanopore_2020,stevens_sequence_2016,campos_high_2021,gao_genome_2021,van_dijk_machine_2021,sun_twenty_2022,kim_application_2020,thudi_genomic_2021,michael_building_2020,shen_omics-based_2022,shahroodi_demeter_2022,
prasad2021soil,Mascher2024,Schreiber2024}

\newcommand{\citecancer}{lawrence_mutational_2013,vogelstein_cancer_2013,ramskold_full-length_2012,baslan_unravelling_2017,shapiro_single-cell_2013,sakamoto_new_2020,jia_high-throughput_2022,lawson_tumour_2018,liu_mrna-based_2023,van_de_sande_applications_2023,chakravarty_clinical_2021,cortes-ciriano_computational_2022,deveson_evaluating_2021,xiao_toward_2021,bolton_cancer_2020,szustakowski_advancing_2021,navin_future_2011,hong_rna_2020,lei_applications_2021,han_single-cell_2022,federici_variants_2020,zhang_singlecell_2021,ren_understanding_2018,tian_cicero_2020,malone_molecular_2020,tang_single-cell_2019,ellsworth_single-cell_2017,zhong_application_2021,stadler_therapeutic_2021,tan_targeted_2022,degasperi_substitution_2022,xu_single-cell_2022,horak_comprehensive_2021,zhang_single-cell_2016,bruno_next_2020,de_luca_fgfr_2020,waarts_targeting_2022,lim_advancing_2020,colomer_when_2020,saadatpour_single-cell_2015,dizman_sequencing_2020,buzdin_rna_2020,xiao_tumor_2021,nandwani_lncrnas_2021,marchetti_error-corrected_2023,chen_next-generation_2021,navin_first_2015}

\newcommand{\citemetagenomicsegs}{hhrlich2011metahit,sunagawa2015structure,fierer2017embracing,danko2021global,paoli2022biosynthetic,edgar2022petabase,prasad2021soil,biodigs2026,Ryon2022,Zhu2025,human2012structure}

\newcommand{\citemgprecision}{kintz2017introducing,dixon2020metagenomics,mousa2024gut,tegegne2025g,virgin2011metagenomics,zhao2024application}

\newcommand{\citemgurgentclinical}{taxt2020rapid,GRUMAZ2020405,sweeney2021rapid,clark2019diagnosis,farnaes2018rapid,gu2021rapid,charalampous2024routine,Heitz2023,Alcolea-Medina2025,LIANG2023101898,Chien2022,Neyton2023,GENG202181,Ren2021}

\newcommand{\citemgbiodiversity}{afshinnekoo2015geospatial,hsu2016urban,sunagawa2015structure,danko2021global,biodigs2026,fierer2017embracing,Zhu2025}

\newcommand{\citemgcommunicable}{john2021next,nagy2021targeted,nieuwenhuijse2017metagenomic,downie2023surveillance}

\newcommand{\citealgoptimization}{zhang2000greedy,slater2005automated,li2018minimap2,myers1999fast,marco2021fast,marcosola2023optimal,grootkoerkamp2024apa2,xin2013accelerating,xin2015shifted,sadasivan2024genomic,tseng2025ultrafast,walia2025ultrafast,kim_fastremap_2022,Ashyralyyev2026gencore,ashyralyyev2026lcpan,alicioglu2024pairwise,manber1993suffix}

\newcommand{\citegraphalgoptimization}{rautiainen2020graphaligner,kim2019hisat2,gao2020abpoa,jain2019pasgal,siren2021pangenomics,Rautiainen2019,Chandra2023,Ivanov2022,Ma2023,Darby2020vargas,Hwang2025MEMO,Romain2023svjedi,Li2020minigraph}

\newcommand{\citehwoptimization}{doblas2025smx,mutlu2023accelerating,alser2022molecules,lou2020helix,lou2018brawl,shahroodi2023swordfish,markus2020benchmarking,subramaniyan2021accelerated,huangfu2018radar,khatamifard2021genvom,gupta2019rapid,li2021pim,angizi2019aligns,zokaee2018aligner,turakhia2018darwin,fujiki2018genax,madhavan2014race,cheng2018bitmapper2,houtgast2018hardware,houtgast2017efficient,zeni2020logan,ahmed2019gasal2,nishimura2017accelerating,de2016cudalign,liu2015gswabe,liu2013cudasw++,liu2009cudasw++,liu2010cudasw++,wilton2015arioc,goyal2017ultra,chen2016spark,chen2014accelerating,chen2021high,fujiki2020seedex,banerjee2018asap,fei2018fpgasw,waidyasooriya2015hardware,chen2015novel,rucci2018swifold,haghi2021fpga,li2021pipebsw,ham2020genesis,ham2021accelerating,wu2019fpga,cali2020genasm,Zhang_2023_alignerD,soysal2025mars,kim2018grim,kaplan2020bioseal,mao2022genpip,dphls2026,wang20202,Walia2024talco,sadasivan2024genomic,Turakhia2025toward,Turakhia2019darwinwga,simon2026processing,eudine2026genpairx,Lindegger2023scrooge}
\newcommand{\citemghwoptimization}{jia2011metabing,kobus2021metacache,wang2023gpmeta,kobus2017accelerating,Su2012,su2013gpumetastorms,Yano2014,saavedra2020mining,zhang2023genomix,cervi2022metagenomic,wu2021sieve,shahroodi2022krakenonmem,shahroodi2022demeter,dashcam23micro,hanhan2022edam,zou2022biohd,dunn2021,shih2023efficient}

\newcommand{\citegraphhwoptimization}{cali2022segram,Zhang2024Harp,Zeng2024asgdp,Shen2024128parallel,Li2024,Mandal2020,Varma2013,Awan2021,Feng2021,Zhang2025,kim2025nmp,Huang2023meg2,Angizi2020Panda,Qiu2017,Zhou2021,Sarkar2021,Varma2017,Varma2016,Goswami2018,Galanos2021,Angizi2020,Sinha2022,Meng2014,Hu2016,Chen2023,Natarajan2018,Ren2018}

\newcommand{\citefilteroptimizationadditional}{singh2021fpga,kim20111,hameed2021alpha,guo2019hardware}

\newcommand{\citegenomiccompression}{chandak2018spring,chandak2017compression,roguski2018fastore,cogo2021genodedup,kokot2022colord,Meng2023,dufort2021renano,kowalski2019pgrc,dufort2020enano,dragenora,yang2025gpufastqlz,chen2023efficient,hach2012scalce,roguski2014dsrc2,Deorowicz2020,lan2021genozip,alyami2019lfastqc,hach2014deez}

\newcommand{\citemapping}{alser2020accelerating, huangfu2018radar, cali2020genasm, turakhia2018darwin, fujiki2018genax, fujiki2020seedex, banerjee2018asap, khatamifard2021genvom, gupta2019rapid, li2021pim, angizi2019aligns, zokaee2018aligner, madhavan2014race, cheng2018bitmapper2, houtgast2018hardware,houtgast2017efficient, goyal2017ultra, chen2016spark, chen2014accelerating, chen2021high, zeni2020logan, ahmed2019gasal2, nishimura2017accelerating, de2016cudalign, liu2015gswabe, liu2013cudasw++, wilton2015arioc, fei2018fpgasw, waidyasooriya2015hardware, chen2015novel, rucci2018swifold, haghi2021fpga, li2021pipebsw, ham2020genesis, ham2021accelerating, wu2019fpga,doblas2025smx,kim2018grim,alser2020sneakysnake,bingol2021gatekeeper,xin2016optimal,mao2022genpip,mansouri2022genstore,xin2015shifted,alser2017gatekeeper,alser2019shouji,alser2017magnet,Zhang_2023_alignerD,cali2022segram,mutlu2023accelerating,alser2022molecules,subramaniyan2021accelerated,xin2013accelerating,kaplan2020bioseal,angizi2020pim}

\newcommand{\citeasm}{cali2020genasm,vsovsic2017edlib,alser2017gatekeeper,alser2017magnet,alser2019shouji,alser2020sneakysnake,kim2018grim,kim2019airlift,kim2024airlifttcbb,needleman1970general,smith1981identification,gotoh1982improved}

\newcommand{\citenongenomegraphisp}{matam2019graphssd,Wang2024ndsearch,lee2022smartsage,Niu2024flashgnn,Lee2024presto,Khadirsharbiyani2024smartgraph,Zhang2025taijigraph,An2023baraddur,Kang2024sting}
\newcommand{\citenongraphgenomeisp}{mansouri2022genstore,abakus23taco,megis,jun2016storage,kim2025nmp,soysal2025mars,zheng2025storage}

\newcommand\tomi[1]{{\color{black}{#1}}}

\definecolor{cornflowerblue}{rgb}{0.39, 0.58, 0.93}
\newcommand\aooo[1]{{\color{black}{#1}}}

\newcommand\mganalysis{metagenomic analysis\xspace}

\newcommand{\circg}[1]{\tikz[baseline=(char.base)]{
           \node[shape=circle,draw=none,inner sep=0.01pt,fill=tealish!100, text=white] (char) {#1};}}

\newcommand{\circr}[1]{\tikz[baseline=(char.base)]{
           \node[shape=circle,draw=none,inner sep=0.01pt,fill=deepred!100, text=white] (char) {#1};}}

\newcommand\nh[1]{{\color{black}{#1}}}
\definecolor{tealish}{RGB}{0,187,161}
\definecolor{deepred}{RGB}{192,0,0}

\setcopyright{none}
\renewcommand\footnotetextcopyrightpermission[1]{} 
\begin{abstract}
    Due to the challenges of analyzing and storing massive volumes of genomic and metagenomic sequence data, \emph{significant} efforts have been made to accelerate (meta)genomic analyses and store sequence data compressed. Despite the benefits of these techniques, we identify two major outstanding problems in accessing stored sequence data and \icsomm{supplying} it to the analysis units: (i) the \icsom{data movement bottleneck due to} moving large amounts of low-reuse data from storage and the unnecessary burden on the rest of the system, and (ii) the \emph{data preparation bottleneck}, where compressed sequence data needs to be first decompressed and formatted before analysis. 

We present \emph{customized storage-centric systems}, which efficiently (i) analyze (meta)genomic data inside storage, and (ii) enable highly-compressed storage and high-performance access of large-scale sequence data, thereby alleviating the overheads of data movement, computation, and data preparation. First, we introduce \emph{GenStore}, an in-storage processing system that filters genomic data not requiring expensive computation directly inside storage. Second, we propose \emph{MegIS}, an in-storage processing system that significantly reduces the data movement overhead of metagenomic analysis. Third, we introduce \emph{GRAINS}, \icsom{a storage-centric system} for analysis on large-scale (meta)genomic graphs in storage. Fourth, we propose \emph{SAGe}, an algorithm-architecture co-design for highly-compressed storage and high-performance access of sequence data. 

We demonstrate that the proposed systems significantly (e.g., by one to two orders of magnitude) improve performance, energy efficiency, and cost-efficiency, \icsom{all at the same time}. We hope these systems facilitate broader adoption of (meta)genomics and inspire research on other data-intensive domains in health and life sciences.
\end{abstract}

\begin{document}

\title{Enabling Fast, Efficient, and Low-Cost\\Genomic and Metagenomic Analyses\\ via Storage-Centric System Designs}

\def\iscacameraready{} %
\newcommand{\hpcapubid}{0000--0000/00\$00.00}

\author{Nika Mansouri Ghiasi \hspace{1em} Onur Mutlu\vspace{1em}}

\affiliation{%
  \institution{SAFARI Research Group}
  \country{ETH Zürich \vspace{0.5em}}
}

\pagestyle{fancy}
\fancyhf{} 
\renewcommand{\headrulewidth}{0pt}

\renewcommand{\headrulewidth}{0pt}

\maketitle

\newcommand{\iscaheight}{0mm}
\ifdefined\eaopen
\renewcommand{\iscaheight}{12mm}
\fi

\setcounter{page}{1}

\vspace{-0.7em}
\section{Genomic and Metagenomic Analyses}

\emph{Genome sequence analysis}, which examines the genomic information of living organisms and other biological entities, plays an important role in many fields, such as tracking outbreaks of communicable diseases~\cite{\citeoutbreak}, personalized and precision medicine~\cite{\citepersonalized}, cancer research~\cite{\citecancer}, pathogen monitoring for food safety~\cite{e002244,TONG2021130}, agriculture~\cite{\citeagriculture}, scientific discovery~\cite{urbanek2018degradation,edgar2022petabase,paoli2022biosynthetic}, and biodiversity conservation~\cite{Hogg2024,lewin2018earth}.
To analyze genomic information computationally, a sample of an organism's or a biological entity's genetic material, typically DNA or RNA that has been reverse-transcribed into DNA~\cite{Houldcroft2017,Jansz2024viral}, undergoes a process called \emph{sequencing}~\cite{\citesequencing}.
Sequencing converts the information from DNA molecules to digital data. Current sequencing technologies \emph{cannot} sequence long DNA molecules end-to-end. Instead, state-of-the-art sequencers generate randomly- and redundantly-sampled smaller and inexact DNA fragments, called \emph{reads}. Sets of genomic reads (called \emph{read sets}) are then used in genomic analysis. Traditional genomics analyzes  sequences of a genomic sample from an individual (or a small group of individuals) of the \emph{same known species}.

Since sometimes a sample contains organisms or biological entities with \emph{different species} present in a \emph{common environment} (e.g., human gut, soil, or oceans), genomic analysis is complemented by \emph{metagenomic analysis}~\cite{\citemetagenomicsegs}. Metagenomic analysis refers to the study of the genome sequences of various organisms \aooo{or biological entities} with different species present in a common environment. Since metagenomics deals with genome sequences whose species are \emph{not known} in advance in many cases, it requires comparisons of the target sequences against large databases of many reference genomes. Metagenomics has led to groundbreaking advances in many fields, such as precision medicine~\cite{\citemgprecision}, urgent clinical settings~\cite{\citemgurgentclinical}, understanding microbial diversity of an environment~\cite{\citemgbiodiversity}, and discovering early warnings of communicable diseases~\cite{\citemgcommunicable}.

As genomic databases grow in complexity, \emph{graph-based (meta)genomic analysis} has emerged as a powerful approach for querying of massive and complex genomic (meta)genomic databases~\cite{Iqbal2012,marchet2021data,karasikov2020metagraph,karasikov2022lossless,karasikov2019sparse,danciu2021topology,fan2023fulgor,bradley2019ultrafast}. 
Sequences in a graph are represented by graph walks~\cite{marchet2021data}. Genome graphs provide two fundamental benefits, making them indispensable, particularly in modern, population-scale genomics~\cite{danko2021global,karasikov2020metagraph,siren2021pangenomics,Sherman2020,taylor2024beyond}.
First, the graph topology, along with its associated metadata,\footnote{Metadata can include the original species or samples of a graph node's sequence~\cite{Iqbal2012,marchet2021data}, associations with patient outcomes~\cite{karasikov2020metagraph}, and more.} offer\icsom{s} vastly \emph{strong expressive power}~\cite{eizenga2020pangenome,Sherman2020,taylor2024beyond,Liao2023}. A graph naturally encodes the evolutionary history and diversity of the organisms in a database, revealing their shared and distinct sequences~\cite{eizenga2020pangenome,Sherman2020,Armstrong2020cactus,bradley2019ultrafast}. This leads to reduced bias and improved analysis accuracy. 
For example, genome graphs improve disease diagnosis by capturing population-specific variants often missed by traditional, linear sequences~\cite{Groza2024,Sherman2020}.
Second, genome graphs leverage the inherent redundancy of genomic data to \emph{avoid redundant computation} on shared sequences. This is because shared genomic regions across database entries are stored only once and do not need to be queried separately.

\section{Prior Work on Improving the Analysis and Storage of (Meta)Genomic Sequence Data}

The adoption of genomic and metagenomic analyses has been rapidly increasing in recent years~\cite{clark2019diagnosis,farnaes2018rapid,sweeney2021rapid,ginsburg2009genomic,chin2011cancer,Ashley2016,bloom2021massively,gilchrist2015whole, dixon2020metagenomics,chiang2019from,chiu2019clinical,kaplan2025pangenomicsbench,das2024systems}, driven by their critical importance and the rapid advancement of sequencing technologies (i.e., reduced costs and increased throughput~\cite{berger2023navigating}). These factors have led to exponential growth in genomic and metagenomic sequence data generation.

Due to the challenges of analyzing and storing massive volumes of genomic and metagenomic data, significant efforts have been made in two directions: \inum{i}~Accelerating (meta)genomic analys\tomi{e}s, and \inum{ii}~Storing (meta)genomic data in compressed forms.

\head{Accelerating Genomic and Metagenomic Analysis} There have been extensive efforts to \tomi{improve the performance and energy efficiency of} genomic and metagenomic analysis. 
For genomics, many works propose efficient heuristics and algorithmic optimizations (e.g.,~\cite{\citealgoptimization,\citegraphalgoptimization,simon2026pim,altschul1990basic}), hardware accelerators (e.g.,~\cite{\citehwoptimization,\citegraphhwoptimization,simon2026pim}), various filters that try to efficiently and accurately prune reads that do not require expensive computation (e.g.,~\cite{alser2020technology,kim2018grim,alser2020accelerating,cali2020genasm,singh2021fpga,nag2019gencache, kim20111, alser2017gatekeeper, alser2017magnet, alser2019shouji, alser2020sneakysnake, bingol2021gatekeeper, hameed2021alpha, guo2019hardware,xin2015shifted,xin2013accelerating}).
For metagenomics, many works propose algorithmic optimizations (e.g.,\tomi{~\cite{lapierre2020metalign,song2024centrifuger,koslicki2016metapalette,Marcelino2020,piro2016dudes,piro2020ganon,pockrandt2022metagenomic,wood2014kraken,karasikov2020metagraph,fan2023fulgor}}), sampling to reduce database size\tomi{s}, \tomi{usually} at the cost of accuracy loss (e.g.,~\cite{kim2016centrifuge,wood2019improved,muller2017metacache,song2024centrifuger,Dilthey2019,Fan2021}), and hardware acceleration (e.g.,~\cite{\citemghwoptimization}).

\head{Storing Genomic and Metagenomic Data in Compressed Forms} It is common practice to store genomic and metagenomic sequence data in compressed forms\tomi{~\cite{berger2023navigating,zhu2013high,Deorowicz2013,giancarlo2013compressive,Betschart2025,Walia2026}} because storing uncompressed sequence data is impractical due to its massive size. In fact, due to the importance of storing sequence data in a space-efficient manner, there exist many compression techniques (e.g.,~\cite{\citegenomiccompression}) specialized for sequence data to achieve significantly higher compression ratios than state-of-the-art general-purpose compression methods (e.g.,~\cite{collet2018zstandard,pavlov20167,Brotli,Katz1991US5051745A,goyal2021dzip,goyal2018deepzip,chen2024ha,bartik2015lz4,liu2018data,fowers2015scalable,chen2021fpga,angerd2022gbdi,gao2024beezip,karandikar2023cdpu,9499902,abali2020data}).

\section{Problem Discussion}
\label{sec:intro.key_problem}

Although there have been \tomi{significant} efforts to improve the analysis and storage of large-scale genomic and metagenomic data, we identify major outstanding problems in accessing stored \tomi{sequence} data and \icsom{supplying} it to the analysis units. \icsom{The first author's PhD thesis\icsomm{~\cite{thesis}} provides a detailed treatment of these problems and proposes solutions to them.}

\head{Overhead of Moving Large Amounts of Low-Reuse Data from the Storage System to Main Memory and Computation Units} Genomic and metagenomic analyses incur unnecessary data movement from the storage system for large amounts of \emph{low-reuse} data. \linebreak
In genomics, while existing filters prune many reads to avoid expensive computation, they still need to first read the entire read set from the storage system \tomi{all the way to the main memory, processor-side caches, register files}, even though a large fraction of the reads would be filtered out and not be reused in the analysis. As detailed in \cite{mansouri2022genstore,arxivGS}, we analyze the state-of-the-art genomic analysis tools and accelerators and find that the movement of large amounts of low-reuse genomic data from the storage system significantly hinders the end-to-end performance and energy-efficiency of genomic analysis due to storage I/O overheads and unnecessary computational burden on the rest of the system.
As we demonstrate in \cite{mansouri2022genstore,arxivGS,thesis},  these overheads can bottleneck the performance \tomi{and energy} of genomic analysis in both conventional (software-based) and emerging (hardware-accelerated) genomics systems, while having a larger impact on systems that reduce other \tomi{(e.g., computation and main memory)} bottlenecks\tomi{~\cite{\citehwoptimization,\citegraphhwoptimization}}.

In metagenomics, analysis also suffers from significant data movement overhead due to the need to access \icsom{very} large amounts of low-reuse data. Since we do not know the species present in a metagenomic sample, \mganalysis requires searching large databases (e.g., several TBs~\cite{ncbi2023,karasikov2020metagraph,shiryev2023indexing,pebblescout,lemane2023kmindex,marchet2023scalable} or more than a hundred TBs in emerging databases~\cite{shiryev2023indexing,pebblescout}) that contain information on different organisms' genomes. Database sizes are expected to increase further in the future, and at a fast pace.\footnote{For example, based on recently published trends, the ENA assembled/annotated sequence database size currently \emph{doubles} every 19.9 months~\cite{enastats}, and the \icsom{BLAST Nucleotide Database (nt)} database size doubled from 2021 to 2022~\cite{ntdouble}.} Two notable reasons for this growth are 1) the rapid evolution of viruses and bacteria~\cite{Lynch2010}, which necessitates frequent updates with new reference genomes\icsom{~\cite{Nasko2018,o2016reference,kim2024airlifttcbb}}, and 2) the fact that databases may include sequences from both highly curated reference genomes and from less curated metagenomic sample sets~\cite{karasikov2020metagraph,shiryev2023indexing}. Particularly, as over 99\% of Earth's microbes remain unidentified and excluded from curated reference genome databases~\cite{jiao2020microbial,Li2024}, the expanded databases improve sensitivity~\cite{Li2024}. Recent advances in the automated and scalable construction of genomic data from more organisms have further contributed to database growth by enabling the rapid addition of new sequences to databases~\cite{rautiainen2023telomere,jarvis2022semi}. 
As we demonstrate in \icsom{\cite{megis,megisarxiv,thesis}}, data movement overhead from the storage system to the rest of the system significantly impacts the end-to-end performance and \tomi{energy efficiency} of \tomi{the state-of-the-art \mganalysis tools}. Due to its low reuse, the data needs to move all the way from the storage system to the main memory, \tomi{processor-side caches, register files,} and \icsom{finally to} processing units for its first use, and it will likely \emph{not} be used again or reused very little during analysis. This unnecessary data movement, combined with the low computation intensity of \mganalysis and the limited I/O (input/output) bandwidth, leads to large storage data movement overheads for \mganalysis. The impact of this overhead becomes even larger on systems that reduce other \tomi{(e.g., computation and main memory)} bottlenecks~\cite{\citemghwoptimization}.

\tomi{As we demonstrate in \cite{grains,grainsarxiv,thesis}, the overhead of moving large amounts of low-reuse data from the storage system and its burden on the rest of the system is also significant in graph-based (meta)genomic analysis. Our evaluation using state-of-the-art tools for graph-based genome analysis shows that this overhead significantly hinders the performance and energy efficiency \icsom{of graph-based genome analysis}.}

\begin{figure}[b]
         \centering
         \includegraphics[width=\columnwidth]{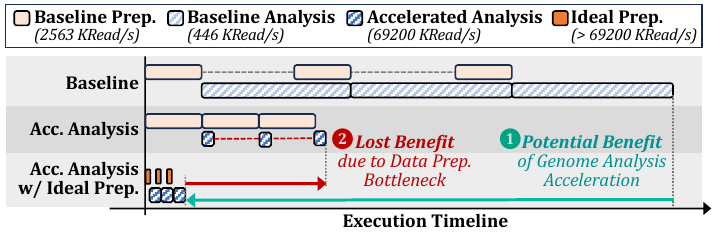}
         \caption{\tomi{Effect of data preparation (i.e., decompressing and formatting genomic sequence data before analysis) on genome analysis performance \icsom{(reproduced from \cite{mansouri2026sage,sagearxiv})}.}}
         \label{fig:intro-motivation}
\end{figure}

\head{Data Preparation Bottleneck} 
\tomi{We introduce and extensively analyze
the \emph{data preparation bottleneck}, where compressed genomic sequence data needs to be first decompressed and formatted before it can be analyzed. As we demonstrate in \cite{mansouri2026sage,sagearxiv,thesis}, the benefits of prior works on accelerating genome sequence analysis \icsom{greatly} diminish due to this bottleneck.
For example, \fig{\ref{fig:intro-motivation}} shows the execution timeline of data preparation and genome analysis for a real-world genomic dataset \icsom{(as detailed in \cite{mansouri2026sage,sagearxiv})} in three different configurations. The evaluated analysis task is read mapping, a fundamental process in genomics. The configurations are 
\inum{i}~\textbf{Baseline:}~a state-of-the-art software analysis tool~\cite{li2018minimap2} with a state-of-the-art software genomic decompressor for data preparation~\cite{chandak2018spring};
\inum{ii}~\textbf{Acc. Analysis:} a state-of-the-art hardware-accelerated analysis tool~\cite{chen2023gem} with the same data preparation \icsom{mechanism} as Baseline;
\inum{iii}~\textbf{Acc. Analysis w/ Ideal Prep.:} the same accelerated analysis with \emph{ideal} \icsom{data} preparation, where \icsom{data} preparation \icsom{time} is \icsom{completely} overlapped with analysis. 
For all configurations, \icsom{data} preparation and \icsom{data} analysis operate in a pipelined manner and in batches (i.e., when decompressing  batch $\#i$, the mapper
analyzes batch $\#i - 1$). Decompressed data batches are directly fed to the analysis stage.
We observe that \circg{1}~hardware acceleration of genome analysis can potentially offer substantial performance benefits; \circr{2}~however, as analysis gets faster, data preparation emerges as a critical bottleneck that hinders the full realization of these benefits. In \cite{mansouri2026sage,sagearxiv,thesis}, we further demonstrate and extensively analyze this bottleneck across various real-world scenarios.}

\section{Goal}

Our goal is to \inum{i}~alleviate the data movement overheads of genomics and metagenomics analyses from the storage system and reduce the computational burden from the rest of the system, and \inum{ii}~mitigate the data preparation bottleneck while achieving high performance, energy efficiency, and compression ratios.
\icsom{To this end, we design lightweight systems such that}
they can seamlessly integrate with a broad range of genomic and metagenomic analysis systems.

\section{Our Approach: Storage-Centric Systems for Genomics and Metagenomics}

The following thesis statement~\cite{thesis} encompasses our approach:

\begin{quote}
\emph{By designing customized storage-centric systems that efficiently}

\emph{1) analyze genomic and metagenomic data inside the storage system}

\emph{2) enable highly-compressed storage and high-performance access of large-scale sequence data,}

\emph{we can alleviate data movement overheads from the storage system, reduce the computational burden from the rest of the system, and mitigate the data preparation bottleneck,}

\emph{thereby significantly \tomi{(e.g., by one to two orders of magnitude)} improving system performance, energy efficiency, and cost of genomic and metagenomic analysis.}
\end{quote}

To alleviate data movement overheads of moving large amount\tomi{s} of low-reuse sequence data from the storage system and reduce the overall \tomi{computational} burden \tomi{from} the rest of the system, we propose \emph{storage-centric computing} (SCC) systems for 1) genomic analysis, 2) metagenomic analysis, and 3) graph-based (meta)genomic analysis. \tomi{SCC refers to processing data inside the storage device, either on the SSD controller (in-storage processing, i.e., ISP) or on the flash dies (in-flash processing, i.e., IFP)\icsom{~\cite{mutlu2025memory,mutlu2022modern,mutlu2019processing,mutlu2019enabling}}. SCC can be a fundamental solution for alleviating this data-movement overhead by processing data where it originally resides\icsom{~\cite{mutlu2025memory,mutlu2022modern,mutlu2019processing,mutlu2019enabling}}.} 
To mitigate the data preparation bottleneck, we propose an algorithm-architecture co-design for highly-compressed storage and high-performance access of sequence data. In the rest of this paper, we briefly overview our new mechanisms developed as part of \icsom{the first author's} dissertation~\cite{thesis}.

\subsection{In-Storage Filters for Genomic Analysis}
\label{sec:gs}

Read mapping~\cite{\citemapping} is a fundamental step in many genomics applications. It is used to identify potential matches and differences between reads of a sequenced genome and an already known genome (called a \emph{reference genome}).
Read mapping is  costly because it needs to perform \emph{approximate string matching (ASM)}~\cite{\citeasm} on large amounts
of data.
To address the computational challenges in genomic analysis, many prior works \tomi{(e.g.,~\cite{\citehwoptimization,\citealgoptimization,\citefilteroptimizationadditional})} propose various approaches such as accurate \emph{filters} that select the reads within a dataset of genomic reads (called a \emph{read set}) that \emph{must} undergo expensive computation. 
While effective at reducing the amount of expensive computation, all such approaches still require the costly movement of large data from storage to the rest of the system, which significantly lower\tomi{s} the end-to-end performance of read mapping in conventional and emerging genomics systems. 

\icsom{We introduce \textbf{GenStore}~\cite{mansouri2022genstore,arxivGS,thesis}}, the first in-storage processing system designed for genomic analysis that significantly reduces both \tomi{storage} data movement and computational overheads of genomic analysis by exploiting low-cost and accurate in-storage filters. 
GenStore leverages hardware/software co-design to address the challenges of in-storage processing,
supporting reads with  1)~different properties such as read lengths and error rates, which highly depend on the sequencing technology, and 2)~different \omcvii{degrees of} genetic variation compared to the reference genome, which highly depends on the genomes that are being compared.
Through rigorous analysis of read mapping processes of reads with different properties and degrees of genetic variation, we meticulously design low-cost hardware accelerators and data/computation flows inside a NAND flash-based solid-state drive (SSD). 

\begin{figure}[b]
    \centering
    \includegraphics[width=\columnwidth]{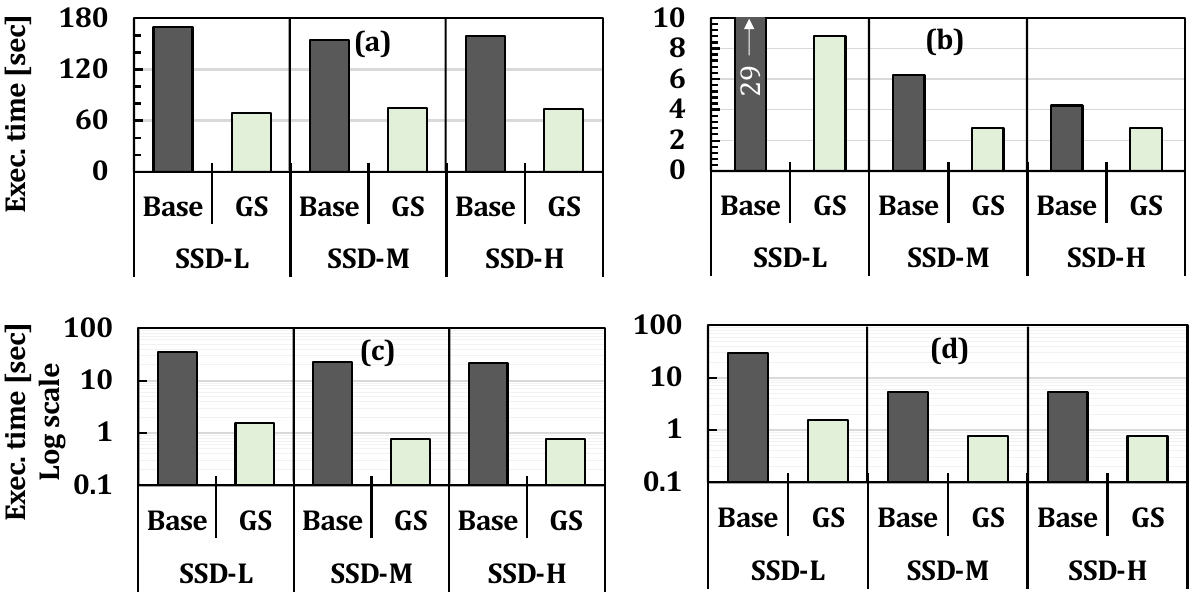}       
    \caption{\icsom{Speedups of GenStore for filtering exactly matching reads \icsomm{in} different SSDs, when integrated with (a)~a software read mapper and (b)~a hardware read mapper, and speedups of GenStore for filtering non-matching reads \icsomm{in} different SSDs, when integrated with (c)~a software read mapper and (d)~a hardware read mapper \icsom{(reproduced from \cite{mansouri2022genstore,arxivGS})}.}}
    \label{fig:gs-perf}
\end{figure}

We evaluate GenStore on a wide range of real genomic datasets (as detailed in \cite{mansouri2022genstore,arxivGS}).
\icsom{\fig{\ref{fig:gs-perf}} shows the benefits of GenStore when integrated with the state-of-the-art software and hardware read mappers. We evaluate the following systems:
\inum{i}~{\textbf{\base:} Minimap2~\cite{li2018minimap2} \omciv{is} a state-of-the-art software read mapper baseline for \jsr{both} short and long reads\omciv{.} GenCache~\cite{nag2019gencache} and Darwin~\cite{turakhia2018darwin} \omciv{are} state-of-the-art hardware read mappers for short and long reads, respectively}; and \inum{ii}~\textbf{\gs:} \base integrated with the {hardware} GenStore filtering accelerators for exactly-matching reads and non-matching reads. GS concurrently filters reads \emph{inside} the SSD while {using} \base to perform read mapping for unfiltered reads. We evaluate each system with three different SSD configurations:
1)~a low-end SSD (\ssdl)~\cite{inteldcs4500} with a SATA3 interface\omciv{~\cite{SATA}}, 
2)~a mid-end SSD (\textsf{SSD-M})~\cite{samsung980pro} using a PCIe Gen3 M.2 interface\omciv{~\cite{PCIE}}, and
3)~a high-end SSD (\textsf{SSD-H})~\cite{samsungPM1735} with a PCIe Gen4 interface\omciv{~\cite{PCIE4}}. \icsommm{Our detailed methodology is described in} \cite{mansouri2022genstore,arxivGS}.
We observe that GenStore significantly improves the read mapping performance of the state-of-the-art software (hardware) baselines.

\icsommm{Our evaluations across various genomic datasets (as detailed in \cite{mansouri2022genstore,arxivGS,thesis})
show that GenStore provides
2.07-6.05$\times$ (\rev{1.52-3.32}$\times$) speedups
over state-of-the-art software (hardware) baselines for \omccc{read sets with high similarity to the reference genome}, and \rev{1.45-33.63}$\times$ (2.70-19.2$\times$) speedups for \omccc{read sets with low similarity to the reference genome}.}}
\tomi{Our analysis shows that GenStore reduces energy consumption by on average (up to) 3.92$\times$ (3.97$\times$) for read sets with high similarity to the reference genome, and on average (up to) 27.17$\times$ (29.25$\times$) for read sets with low similarity to the reference genome.}

\tomi{GenStore is published at the International Conference on Architectural Support for Programming Languages and Operating Systems (ASPLOS) in 2022~\cite{mansouri2022genstore}. It is fully open-sourced at~\cite{gssource}. An extended version of the ASPLOS 2022 paper is available on arXiv~\cite{arxivGS}. A significant amount of work \icsomm{(e.g.,~\cite{soysal2025mars,abakus23taco,zheng2025storage,kabra2025ciphermatch,chen2025reis,megis,grains,mansouri2026sage})} has already been influenced by GenStore, as we discuss in \cite{thesis}.}

\vspace{-0.5em}
\subsection{Cooperative In-Storage Processing for Metagenomic Analysis}

In-storage processing can be a fundamental solution for 
reducing the overhead of moving large amounts of low-reuse data from the storage system and its burden on the rest of the system.
However, designing an in-storage processing system for metagenomics is challenging because \tomi{none of the existing tools for} \mganalysis can be directly implemented in storage effectively due to the hardware limitations of modern SSDs.

\icsom{We propose} \textbf{MegIS}~\cite{megis,megisarxiv,thesis}, the \emph{first} in-storage processing system designed to significantly reduce the data movement overhead of the end-to-end metagenomic analysis pipeline. 
\tomi{The key idea of MegIS is to enable  \emph{cooperative} ISP for metagenomics, where we do not solely focus on processing inside the storage system but, instead, we capitalize on the strengths of processing \emph{both inside and outside} the storage system. We enable cooperative ISP via a synergistic hardware/software co-design between the storage system and the host system.
We design MegIS as an efficient pipeline between the SSD and the host system to \inum{i}~\emph{leverage} and \inum{ii}~\emph{orchestrate} the capabilities of both. Based on our rigorous analysis of the end-to-end \mganalysis pipeline, we propose a new hardware/software co-designed accelerator framework that consists of five aspects.
First, we partition and map different parts of the \mganalysis pipeline to the host and the ISP system such that each part is executed on the most suitable architecture. 
Second, we coordinate the data/computation flow between the host and the SSD such that MegIS \inum{i}~completely overlaps the data transfer time between them with computation time to reduce the communication overhead, \inum{ii}~leverages SSD bandwidth efficiently, and \inum{iii}~does not require large DRAM inside the SSD or a large number of writes to the flash chips. 
Third, we devise storage technology-aware metagenomics algorithm optimizations to enable efficient access patterns to the SSD. 
Fourth, we design lightweight in-storage accelerators to perform MegIS's ISP functionalities while minimizing the required SRAM/DRAM buffer spaces inside the SSD. 
Fifth, we design an efficient data mapping scheme and Flash Translation Layer (FTL) specialized to
the characteristics of \mganalysis to leverage the SSD's full internal bandwidth.} 
MegIS's design is flexible, capable of supporting different types of metagenomic datasets, and can be integrated into various metagenomic analysis pipelines. 

Our evaluation shows that MegIS outperforms \tomi{two} state-of-the-art performance- and accuracy-optimized software metagenomic tools\tomi{~\cite{wood2019improved,lapierre2020metalign}} by 2.7$\times$--37.2$\times$ and 6.9$\times$--100.2$\times$, respectively,  while matching the accuracy of the accuracy-optimized tool. MegIS \icsommm{provides} 1.5$\times$--5.1$\times$ speedup compared to the state-of-the-art metagenomic hardware-accelerated  (using processing-in-memory~\cite{wu2021sieve}) tool, while achieving significantly higher accuracy. \tomi{MegIS provides large average energy reductions
of 5.4$\times$ and 1.9$\times$ compared to software and hardware performance-optimized baselines, and 15.2$\times$ compared to the accuracy-optimized baseline.}

\newcommand\poptp{\textsf{P-Opt\_\icsomm{\$\$\$}}\xspace}
\newcommand\aoptp{\textsf{A-Opt\_\icsomm{\$\$\$}}\xspace}
\newcommand\poptc{\textsf{P-Opt\_\icsomm{\$}}\xspace}
\newcommand\aoptc{\textsf{A-Opt\_\icsomm{\$}}\xspace}
\newcommand\msc{\textsf{MS\_\icsomm{\$}}\xspace}

\begin{figure}[h]
\centering
 \vspace{1em}
 \includegraphics[width=\linewidth]{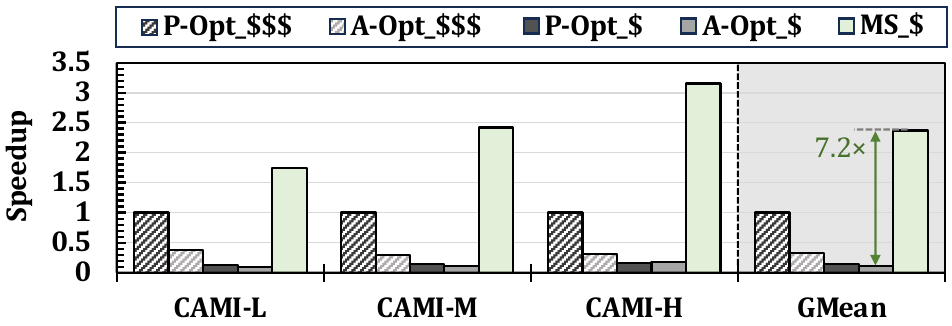}
\caption{Speedup of MegIS on a cost-optimized system over baselines on \omiii{cost-/}performance-optimized systems \icsom{(reproduced from \cite{megis,megisarxiv})}.}
\vspace{1em}
\label{fig:ms-ce}
\end{figure}

\icsom{\icsommm{Importantly}, MegIS also improves system cost-efficiency. \icsommm{This is} because \icsommm{performing ISP alleviates complex hardware needs in the rest of the system and thus reduces cost versus a processor-centric analysis system}.
\fig{\ref{fig:ms-ce}}  compares MegIS on a cost-optimized system with a cost-optimized SSD~\cite{samsung870evo} and 64-GB host DRAM (\msc) to a state-of-the-art performance optimized baseline (P-Opt)~\cite{wood_improved_2019} and a \icsomm{state-of-the-art} accuracy-optimized baseline (A-Opt)~\cite{lapierre2020metalign} 1) on the same system (\poptc and \aoptc) and 2) on a performance-optimized system with a performance-optimized SSD~\cite{samsungPM1735} and 1-TB host DRAM (\poptp and \aoptp).\footnote{For the performance-optimized system, we calculate the cost of 1TB DRAM to be roughly \icsommm{30,100} USD (8$\times$ 128GB modules~\cite{samsung8GBDDR4}) and the cost of the performance-optimized SSD to be roughly 875 USD. For the cost-optimized system, we calculate the cost of 64GB DRAM to be roughly \icsomm{614} USD (8$\times$ 8GB modules~\cite{samsung128GBDDR4}, assuming the same number of memory channels as the performance-optimized system) and the cost of the cost-optimized SSD to be roughly 346 USD.} \icsommm{Our detailed methodology is described in} \cite{megis,megisarxiv,thesis}.

We make two key observations. First, MegIS on the cost-optimized system outperforms the baselines even when they run on the performance-optimized system. \msc provides 2.4$\times$ and 7.2$\times$ average speedup compared to \poptp and \aoptp, respectively. Note that \msc provides the same accuracy as \aoptp and significantly higher accuracy than \poptp. 
Second, baselines on the cost-optimized system experience significantly worse performance compared to when they run on the performance-optimized system. \poptc leads to 6.8$\times$ (7.7$\times$) average (maximum) slowdown over \poptp, and \aoptc leads to 2.8$\times$ (4.2$\times$) average (maximum) slowdown over \aoptp. 

We conclude that MegIS improves system cost-efficiency, while providing high performance and accuracy.
This is critical to \icsommm{enabling \inum{i}~wider adoption} of metagenomic analysis and \icsommm{\inum{ii}~analysis on low-cost, portable devices,} which is increasingly important due to the advances of compact portable sequencers~\cite{minion21,jain2016oxford,cali2017nanopore} for on-site metagenomics~\cite{pomerantz2018real, chiang2019from,mutlu2023accelerating,alser2022molecules}.}

\tomi{MegIS is published at the International Symposium on Computer Architecture (ISCA) in 2024~\cite{megis}. It is fully open-sourced at~\cite{megissource}. An extended version of the ISCA 2024 paper is available on arXiv~\cite{megisarxiv}. MegIS has already influenced several subsequent works \icsomm{(e.g.,~\cite{grains,mansouri2026sage,sagearxiv,chen2025reis,kabra2025ciphermatch})}, as we discuss in \cite{thesis}.}

\subsection{Storage-Aware Algorithm-Architecture Co-Design for Graph-Based Genomic and Metagenomic Analys\tomi{e}s}
\vspace{0.7em}
 
Processing data directly inside the storage device can be a fundamental solution for mitigating storage I/O overheads\icsom{~\cite{mutlu2025memory,mutlu2022modern,mutlu2019processing,mutlu2019enabling}}. However, none of the existing tools for graph-based (meta)genomic analys\tomi{e}s \tomi{(e.g.,~\cite{\citegraphhwoptimization,\citegraphalgoptimization})} can be efficiently \tomi{implemented} inside the storage system due to the limited internal hardware resources in modern SSDs. At the same time, prior storage-centric systems developed for \inum{i}~traditional, linear non-graph-based genomic and metagenomic analys\tomi{e}s \tomi{(e.g.,~\cite{\citenongraphgenomeisp})} or \inum{ii}~conventional, non-genomic graph analysis \tomi{(e.g.,~\cite{\citenongenomegraphisp})} are unsuitable for the unique data structures and access patterns of graph-based genomic and metagenomic analysis.

\icsom{We introduce} \textbf{GRAINS}~\cite{grains,grainsarxiv,thesis}, the \emph{first} system for analysis \tomi{on} large-scale \underline{\textbf{g}}enomic and metagenomic sequence g\underline{\textbf{ra}}phs \underline{\textbf{in}} \underline{\textbf{s}}torage. 
Through our detailed examination of typical analysis pipelines on large-scale sequence graphs, we perform storage-aware algorithm\tomi{-}architecture co-design to \inum{i}~make the pipelines more storage-friendly and \inum{ii}~improve performance, energy-efficiency, and cost-effectiveness via in-storage and in-flash processing. 
GRAINS's co-design is based on three key aspects.
First, we propose a new batching technique and execution flow, based on unique features of sequence graphs, that reduces the number of random \tomi{storage} accesses. Second, via in-flash and in-storage processing, we avoid transferring low-reuse or unused flash pages, preventing SSD channel and external I/O bandwidth waste. Third, to enable leveraging the full parallelism of all flash dies during in-flash processing, we design an effective, yet lightweight, scheduling technique, enabled by re-purposing the existing SSD structures in a new way.
GRAINS's design is versatile and flexible as it supports \tomi{major} operations on sequence graphs and can be integrated in various analysis pipelines. 

Our evaluation shows that GRAINS provides 2.7$\times$–47.8$\times$ speedup over the state-of-the-art software baselines\tomi{~\cite{karasikov2020metagraph,karasikov2022lossless,danciu2021topology,karasikov2019sparse}}, and 1.5$\times$–17.0$\times$ speedup over a hardware-accelerated baseline. \tomi{GRAINS provides significantly higher energy efficiency of on average 16.4$\times$ and 9.8$\times$ over software and hardware baselines, respectively.}

\newcommand\fg{\textsf{FG}\xspace}
\newcommand\mg{\textsf{MG}\xspace}
\newcommand\pim{\textsf{IdealAccMem}\xspace}
\newcommand\grn{\textsf{GRN}\xspace}
\newcommand\grnext{\textsf{GRN-Ext}\xspace}

\begin{figure}[b]
    \centering
    \includegraphics[width=\columnwidth]{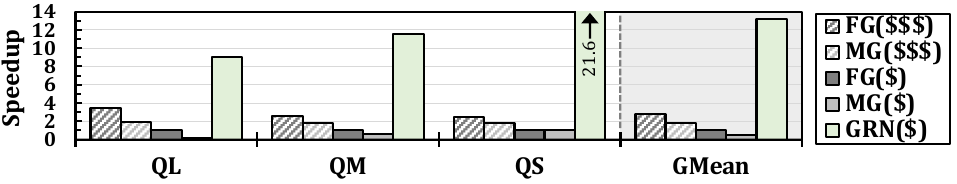}       
    \caption{Speedup of GRAINS on a low-cost system over baselines on low-cost and performance-optimized systems \icsom{(reproduced from \cite{grains,grainsarxiv})}.}
    \label{fig:grn-eval-cost}
\end{figure}

\icsom{GRAINS also improves system cost-efficiency since \icsommm{performing ISP alleviates complex hardware needs in the rest of the system and thus reduces cost versus a processor-centric analysis system.}
We show this by evaluating two systems: a cost-optimized system (\$) with 64-GB DRAM and an SSD with a PCIe Gen4 interface~\cite{samsungPM1735}, and a costlier, performance-optimized system (\$\$\$) with 1.5-TB DRAM and an SSD with a PCIe Gen5 interface~\cite{samsung9100PRO}. \icsommm{More on our methodology is in}~\cite{grains,grainsarxiv,thesis}. \fig{\ref{fig:grn-eval-cost}} compares GRAINS (\grn{}) on the cost-optimized system with the baselines (\fg{}~\cite{fan2023fulgor} and \mg{}~\cite{karasikov2020metagraph,karasikov2022lossless,danciu2021topology,karasikov2019sparse}) on both systems. To fairly evaluate the baselines when DRAM is smaller than the graph \icsomm{size}, we reduce I/O overheads as much as possible in software for this scenario. To this end, 
we partition the graph such that each subgraph fits in the host DRAM, so random accesses to the graph do not repeatedly access the SSD. 
We partition the graph using 
database entry metadata
to guide partitioning choices, 
placing entries 
that likely share many k-mers
into the same partition.
Despite the benefits of this optimization, two sources of overhead remain: \inum{i}~the I/O overhead of transferring subgraphs,
and \inum{ii}~the need to query against each subgraph.

We make two observations.
First, GRAINS on the cost-optimized system significantly outperforms the baselines even on the performance-optimized system. \grn{}\_\$ provides 4.7$\times$ and 5.2$\times$ average speedup over \fg{}\_\$\$\$ and \mg{}\_\$\$\$, respectively. 
Second, \icsommm{the speedups of GRAINS over the baselines increase further when the baselines also run on the cost-optimized system.}
When compared on the same cost-optimized system, \grn{}\_\$ provides 13.2$\times$ and 26.9$\times$ average speedup over \fg{} and \mg{}, respectively.

We conclude that GRAINS improves system cost-efficiency and performance, which is critical for facilitating \inum{i}~the wide adoption of graph-based genome analysis, 
and \inum{ii}~analysis on low-cost, portable devices, a scenario rising in importance with the development of portable sequencing devices~\cite{MinIONMk1CO,palatnick2020igenomics,Ballard2018,Oehler2023} for on-site genome analysis~\cite{pomerantz2018real, chiang2019from,mutlu2023accelerating,alser2022molecules}. 
}

\icsom{GRAINS is published at the International Symposium on Computer Architecture (ISCA) in 2026~\cite{grains}. It will be fully open-sourced at~\cite{grainsgithub}. An extended version is available on \icsomm{arXiv}~\cite{grainsarxiv}.}

\subsection{Algorithm-Architecture Co-Design for Highly-Compressed Storage and High-Performance Access of Sequence Data}
\vspace{0.7em}

\tomi{Given the importance of genomics and the exponentially growing volumes of sequence data, there are extensive efforts to accelerate genomic analysis (e.g.,~\cite{\citehwoptimization}). We demonstrate a major bottleneck that significantly limits and diminishes the benefits of state-of-the-art genomic analysis accelerators~\cite{mansouri2026sage,sagearxiv,thesis}: the data preparation bottleneck, where genomic sequence data is stored in compressed form and needs to be \emph{first} decompressed and formatted before an accelerator can operate on it.} 

To mitigate the data preparation bottleneck, \icsom{we propose} \textbf{SAGe}~\cite{mansouri2026sage,sagearxiv,thesis}, an algorithm-architecture co-design for highly-compressed \textbf{\underline{s}}torage and high-performance \textbf{\underline{a}}ccess of large-scale \textbf{\underline{ge}}nomic sequence data. The key challenge is to improve data preparation performance while maintaining high compression ratios (comparable to genomic-specific compression algorithms) at low hardware cost. 
\tomi{SAGe addresses this challenge based on the \textbf{key insight} that the information encoded in genomic data follows specific trends, shaped by factors such as sequencing technology (e.g., error rates and read lengths) and common genetic phenomena (e.g., typical spatial distributions of genetic variations within genomes). By carefully exploiting these characteristics to synergistically co-design algorithms and hardware, SAGe achieves high compression ratios comparable to state-of-the-art genomic compressors, while enabling low decompression latencies, using only lightweight hardware and efficient streaming accesses. SAGe's co-design consists of new}
\inum{i}~lossless (de)compression algorithm\icsom{s}, \inum{ii}~hardware \nh{that decompresses data with lightweight operations and efficient streaming accesses}, \inum{iii}~storage data layout, and \inum{iv}~interface commands to access data. SAGe is highly versatile, as it supports datasets from different sequencing technologies and species.
Due to its lightweight design, SAGe can be seamlessly integrated with a broad range of hardware accelerators for genome sequence analysis to mitigate their data preparation bottlenecks.

\newcommand\pigz{\texttt{pigz}\xspace}
\newcommand\nspring{\texttt{(N)Spr}\xspace}
\newcommand\nsacc{\texttt{(N)SprAC}\xspace}
\newcommand\idec{\texttt{0TimeDec}\xspace}
\newcommand\psw{\texttt{SAGeSW}\xspace}
\newcommand\pmain{\texttt{SAGe}\xspace}
\newcommand\ssdgi{\texttt{SAGe\textsubscript{SSD}}\xspace}
\newcommand\map{\texttt{Map}\xspace}
\newcommand\isfmap{\texttt{ISF}\xspace}

\icsom{We evaluate the end-to-end performance of various genome analysis systems, where execution includes \emph{both} data preparation and genome analysis.
For \textbf{data preparation}, we use
\inum{i}~\pigz: A parallel version~\cite{adler2015pigz} of gzip, a commonly-used general compressor;
\inum{ii}~\nspring: Spring~\cite{chandak2018spring} and NanoSpring~\cite{Meng2023}, state-of-the-art compressors for short and long reads, respectively. \pigz and \nspring run on a high-end system, as detailed in \cite{mansouri2026sage,sagearxiv}. 
\inum{iii}~\nsacc: \nspring integrated with an idealized BWT-accelerator, as detailed in \cite{mansouri2026sage,sagearxiv};
\inum{iv}~\idec: an idealized decompressor (with zero decompression time), but inefficient for integration in resource-constrained environments (e.g., in our evaluations, for integration with a near-data processing genome sequence analysis system);
\inum{v}~\psw: SAGe with its decompression in software, running on the host system (i.e, the same high-end system used for \pigz and \nspring);
\inum{vi}~\pmain: SAGe's full implementation, with its decompression in hardware;
and \inum{vii}~\ssdgi: \pmain with its hardware implemented in the SSD, to integrate with an NDP genome analysis system on the same chip. 
For \textbf{genome sequence analysis}, we integrate all data preparation configurations with a state-of-the-art read mapping accelerator, GEM~\cite{chen2023gem}. To show SAGe's suitability \icsommm{to} resource-constrained environments, we evaluate SAGe{}'s implementation inside the SSD to integrate with a state-of-the-art NDP genome analysis system, GenStore~\cite{mansouri2022genstore,arxivGS} (\sect{\ref{sec:gs}}).
The resulting pipeline performs data preparation $\rightarrow$ GenStore's in-storage filtering (ISF) $\rightarrow$ mapping. The benefits of this pipeline arise from both SAGe and the ISF \cite{mansouri2022genstore}. The key to realizing this pipeline is that SAGe is the only data preparation configuration that is lightweight enough for efficient implementation inside the SSD. Without SAGe, ISF would require genomic data to be stored uncompressed, which is inefficient, or to decompress data outside SSD, which undermines the fundamental benefits of NDP.   
Further information on \icsommm{our} evaluation methodology \icsommm{is in} \cite{mansouri2026sage,sagearxiv}.}

\begin{figure*}[b]
  \centering
    \includegraphics[width=2\columnwidth]{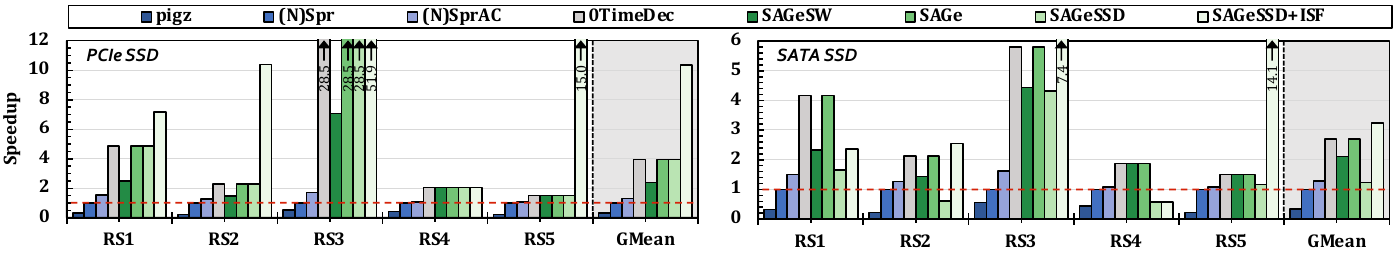}
  \caption{End-to-end speedup for different read sets \icsom{(reproduced from \cite{mansouri2026sage,sagearxiv})}.}
  \label{fig:eval-new-full} 
\end{figure*}

\icsom{\fig{\ref{fig:eval-new-full}} shows end-to-end performance normalized to \nspring.
We make five key observations. 
First, SAGe provides significant speedups. On the system with the PCIe (SATA) SSD, \pmain{} leads to 12.3$\times$ (8.1$\times$),  3.9$\times$ (2.7$\times$), and 3.0$\times$ (2.1$\times$) average speedup over \pigz{}, \nspring, and \nsacc, respectively. 
Second, \pmain{} matches \idec{} in performance because \pmain{} fully hides the decompression overhead in the execution pipeline (I/O, decompression, and read mapping are pipelined, so overall throughput depends on the slowest stage). 
Third, due to its efficient access patterns, SAGe's implementation in software (\psw) also leads to speedups (2.3$\times$ on average) over \nspring{}. However, \psw's decompression still bottlenecks end-to-end performance and leads to up to 4.0$\times$ slowdown over \pmain. 
Fourth, by efficiently integrating \ssdgi{} with \isfmap{} (i.e., the in-storage filter implemented inside the resource-constrained environment of the SSD), \ssdgi{}+\isfmap leads to 7.8$\times$ (2.5$\times$) average speedups over \nsacc{} on the system with the PCIe (SATA) SSD. 
\ssdgi{}+\isfmap outperforms \pmain{} in all cases, except when \inum{i}~the input and application do not largely take advantage of \isfmap{} and \inum{ii}~the SSD's limited external bandwidth bottlenecks performance (e.g., RS1 and RS4 with the SATA SSD). In these cases, the \pmain{} configuration should be used to decompress data outside the SSD to avoid moving larger decompressed data through the limited-bandwidth storage interface.
Fifth, regardless of how \icsomm{optimized} decompression tools are for performance, if their high resource requirements make them unsuitable for adoption in resource-constrained environments, they miss out on the benefits of a wide range of genome analysis systems that are implemented in such constrained environments. For example, \idec{} cannot efficiently and cost-effectively integrate with \isfmap{} implemented inside the SSD and, as shown in our evaluations, it ends up being on average 1.8$\times$ (up to  9.9$\times$) slower than \ssdgi{}+\isfmap.}

\icsommm{In addition to performance benefits,} SAGe also improves the average end-to-end energy efficiency of two state-of-the-art genome sequence analysis accelerators\tomi{~\cite{chen2023gem,mansouri2022genstore,arxivGS}} by  \nh{13.0$\times$--34.0$\times$} compared to when the accelerators rely on state-of-the-art software and hardware decompression tools.

\tomi{SAGe is published at the IEEE International Symposium on High-Performance Computer Architecture (HPCA) in 2026~\cite{mansouri2026sage}. We plan to open-source SAGe to facilitate future research. An extended version of the HPCA 2026 paper is available on arXiv~\cite{sagearxiv}.}

\section{Concluding Remarks} 
\vspace{0.7em}
We identify that although there have been significant efforts to improve the analysis and storage of large-scale genomic and metagenomic data, there are major outstanding problems in accessing stored sequence data and feeding it to the analysis units. These problems arise from \inum{i} the overhead of moving large amounts of low-reuse data from the storage system and the unnecessary burden on the rest of the system (e.g., main memory and computation units), and \inum{ii} the data preparation bottleneck, where compressed sequence data needs to be first decompressed and formatted before it can be analyzed. To alleviate data movement overheads and reduce the overall computational burden of low-reuse data,
we introduce three new storage-centric computing systems for (meta)genomics: GenStore~\cite{mansouri2022genstore,arxivGS}, MegIS~\cite{megis,megisarxiv}, and GRAINS~\cite{grains,grainsarxiv}. To mitigate
the data preparation bottleneck, we propose SAGe~\cite{mansouri2026sage,sagearxiv} an algorithm-architecture co-design for highly-compressed storage and high-performance access of sequence data. We demonstrate that the proposed systems significantly improve system performance, energy efficiency, and cost-efficiency of (meta)genomic analysis. We hope that the storage-centric systems proposed in this \icsom{work~\cite{thesis}} facilitate the broader adoption of (meta)genomic analyses and inspire future research to fundamentally improve the performance, energy efficiency, and cost-effectiveness of other data-intensive application domains related to health and life sciences.

\begin{acks}
\vspace{0.7em}

\icsom{We thank the SAFARI group members for feedback and the stimulating, inclusive, intellectual, and scientific environment. We acknowledge the generous gifts and support provided by our industrial partners, including Google, Huawei, Intel, Microsoft, and VMware. This research was partially supported by the European Union’s Horizon Program for research and innovation under Grant  No. 101047160 (project BioPIM), the Swiss National Science Foundation (SNSF), Semiconductor Research Corporation (SRC), the ETH Future Computing Laboratory (EFCL), Huawei ZRC Storage Team, and the AI Chip Center for Emerging Smart Systems Limited (ACCESS). No AI or LLM help was used in creating this work.} 

\end{acks}

\bibliographystyle{ACM-Reference-Format}
\bibliography{refs}

\end{document}